%% file: DFE_THP.tex
\documentclass{article}
\usepackage{spconf,psfrag,amsmath,epsfig,amssymb,array,setspace}
\input{def.tex}
\newtheorem{lem}{Lemma}

\input{DFE_THP_title.tex}
\begin{document}
\sloppy
\ninept
\maketitle
\input{DFE_THP_abstract.tex}
\input{DFE_THP_intro.tex}
\input{DFE_THP_model.tex}
\input{DFE_THP_design1.tex}
\input{DFE_THP_design2.tex}
\input{DFE_THP_design3.tex}
\input{DFE_THP_sim.tex}
\input{DFE_THP_conc.tex}

\small
\singlespacing
\bibliographystyle{IEEEbib}
\bibliography{../References/IEEEabrv,../References/ref,../References/ref_THP,../References/QRS,../References/DFE,../References/LinPrecoding,../References/Capacity}
\end{document}

%% file: def.tex
\def \Diag  	{\operatorname{Diag}}
\def \tr  	{\textrm{tr}}

\def \MSE 	{\mathbf{E}}
\def \majx      {\prec_{\times}}

\def \I		{{\mathbf I}}

\def \H		{{\mathbf H}}
\def \E		{{\mathbf E}}

\def \G		{{\mathbf G}}
\def \P		{{\mathbf P}}
\def \A		{{\mathbf A}}
\def \B		{{\mathbf B}}
\def \C		{{\mathbf C}}

\def \V		{{\mathbf V}}
\def \U		{{\mathbf U}}
\def \M		{{\mathbf M}}

\def \R		{{\mathbf R}}

\def \L		{{\mathbf L}}
\def \R		{{\mathbf R}}
\def \Q		{{\mathbf Q}}

\def \Lambdab	{{{\mathbf \Lambda}}}
\def \Phib	{{\mathbf \Phi}}

\def \Ptot 	{P_{\textrm{total}}}
\def \Lff  	{\mathbf{L}_{1 1}}
\def \Lkk  	{\mathbf{L}_{K K}}
\def \Lii  	{\mathbf{L}_{i i}}
\def \MSEii  	{{\MSE_{i i}}}
\def \Lall 	{{\mathbf{L}_{1 1}}, \dots,{\mathbf{L}_{K K}}}

\def \sigv 	{\sigma_{v}^{2}}
\def \sig  	{\sigma}

\def \0		{{\mathbf {0}}}
\def \e		{{\mathbf e}}
\def \s		{{\mathbf s}}

\def \e		{{\mathbf e}}
\def \x		{{\mathbf x}} 
\def \y		{{\mathbf y}}
\def \n		{{\mathbf n}}
\def \b		{{\mathbf b}}
\def \u		{{\mathbf u}}

\def \e		{{\mathbf e}}
\def \v		{{\mathbf v}}
\def \u		{{\mathbf u}}
\def \i		{{\mathbf i}}
\def \a		{{\mathbf a}}
\def \b		{{\mathbf b}}
\def \g		{{\mathbf g}}
\def \m		{{\mathbf m}}

\def \l		{{\boldsymbol{\mit{l}}}}
\def \lb 	{\overline{ \l }}

\def \Q		{\mathbf {Q}}

\def \uh 	{\hat{\mathbf{u}}}
\def \sh 	{\hat{\mathbf{s}}}

\def \Kh 	{\hat{K}}

%% file: DFE_THP_title.tex
\title{A Framework for Designing MIMO systems with Decision Feedback Equalization or Tomlinson-Harashima Precoding}
\name{M. Botros Shenouda and T. N. Davidson
\thanks{
This work was supported in part by Natural Sciences and Engineering Research Council of Canada. 
The work of the second author is also supported by the Canada Research Chairs Program.
}
}
\address{Department of Electrical and Computer Engineering\\
McMaster University,  Hamilton, Ontario, L8S4K1, Canada}

%% file: DFE_THP_abstract.tex
\begin{abstract}
We consider joint transceiver design for  general Multiple-Input Multiple-Output  communication systems  that implement interference (pre-)subtraction, such as those based on Decision Feedback Equalization (DFE) or Tomlinson-Harashima precoding (THP).
We develop a unified framework for joint transceiver design by considering design criteria that are expressed as functions of the Mean Square Error (MSE) of the individual data streams. 
By deriving two inequalities that involve the logarithms of the individual MSEs, we obtain optimal designs for two classes of communication objectives, namely those that are  Schur-convex and  Schur-concave functions of these logarithms.
For Schur-convex objectives, the optimal design results in data streams with equal MSEs. 
This design simultaneously minimizes the total MSE and maximizes the mutual information for the DFE-based model.
For Schur-concave objectives, the optimal  DFE  design results in linear equalization and the optimal THP design results in linear precoding. 
The proposed framework embraces a wide range of design objectives and can be regarded as a counterpart of the existing framework of linear transceiver  design.

\end{abstract}

%% file: DFE_THP_intro.tex
\section{Introduction} 
One of the key advantages of Multiple-Input Multiple-Output (MIMO) communications schemes is that they facilitate the simultaneous transmission of multiple data streams.
Typically, such schemes involve processing of the data streams at the transmitter (precoding) and processing of the received signals (equalization) to ``match"  the transmission to the channel and to mitigate the interference between the received streams at reasonable computational cost.
One approach to the design of such a scheme is to focus on linear precoding and linear equalization; e.g. \cite{Palomar_2003, Scaglione_1999a}. 
An alternative approach that offers some advantages is to allow interference (pre-)subtraction at either the transmitter or the receiver. 
This approach includes schemes with linear precoding and Decision Feedback Equalization (DFE), and schemes with Tomlinson-Harashima precoding (THP) and linear equalization, and will be the focus of this paper.

A large number of design strategies have been proposed for the class of linear MIMO transceivers (e.g., \cite{Scaglione_1999a}), and a uniform framework that encompasses many of these designs was proposed in \cite{Palomar_2003}. 
This framework consists of functions that capture a broad range of communication objectives, namely those  that are  Schur-convex and Schur-concave functions of the mean square error (MSE) of each data stream. 
For the class of interference (pre-)subtraction, designs  for DFE based schemes using an MMSE criterion receiver were considered in \cite{THP_Simone, Xu_2006_DFE}, and designs  subject to a zero-forcing constraint were considered in \cite{Jiang_2005_GMD, Zhang_2005_QRS}. 
Some THP counterparts of these designs were presented in \cite{THP_Simone} and \cite{F_THP_Book}, respectively.

In this paper, we develop a broadly applicable framework for joint transmitter and receiver design for MIMO systems with a DFE or a THP. 
We consider the broad range of design criteria that can be expressed as either Schur-convex or Schur-concave functions of the logarithm of the MSE of each data stream, and we provide optimal transceiver designs for these two classes.
In addition to providing a generalization of existing designs based on the overall MSE, these classes of functions embrace other design criteria such as minimizing the maximum of the individual MSEs, or minimizing a  weighted geometric mean of the MSEs. 
Moreover, for the DFE model, design criteria expressed in terms of the signal to interference-plus-noise ratio (SINR) and bit error rate (BER) of each stream are  included in the set of objectives covered by these classes.
Interestingly, the optimal design for both Schur-convex and  Schur-convex  objectives yields a diagonal MSE matrix.
For Schur-convex objectives, the optimal design results in data streams with equal MSEs. 
Furthermore, for the DFE model, the optimal design for this class simultaneously minimizes the total MSE and  maximizes the mutual information. 
For Schur-concave objectives, the optimal design results in linear precoding and equalization.
From a boarder prospective, the proposed framework can be viewed as a counterpart  for the design of DFE-based and THP-based transceivers of the unified framework for the design of linear transceivers in \cite{Palomar_2003}.

%% file: DFE_THP_model.tex
\section{Two System Models}  \label{DFE_THP_sec:model}
\begin{figure}
\centering
\psfrag{h}{$\hat{\tiny{\mathbf{s}}}$}
\psfrag{s}{$\tiny{\mathbf{s}}$}
\psfrag{x}{$\tiny{\mathbf{x}}$}
\psfrag{y}{$\tiny{\mathbf{y}}$}
\includegraphics[width=0.34\textwidth]{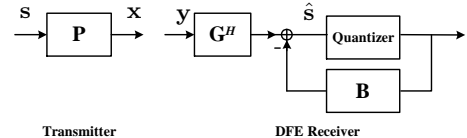}
\caption{MIMO transceiver with Decision Feedback Equalization.}
\label{DFE}
\end{figure}
\begin{figure}
\centering
\includegraphics[width=0.28\textwidth]{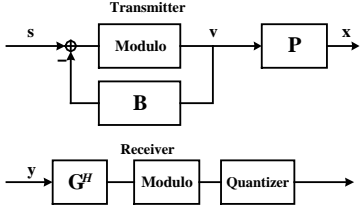}
\caption{ MIMO transceiver with Tomlinson-Harashima precoding}
\label{THP1}
\end{figure}

We consider a generic  MIMO communication system in which the received signal can be written as $\y = \H \x +  \n $, where $\H \in \mathbb{C}^{N_r \times N_t}$  represents the channel, the transmitted vector $\x$ is synthesized from a vector $\s \in \mathbb{C}^{K}$ of data symbols,  and the additive noise has zero-mean and covariance matrix $\mathrm{E}_n \{ \n \n^H\} = \R_n$.
We will consider a general design approach that encompasses several design criteria for two communication systems,
namely those systems with  linear precoding at the transmitter and a DFE at the receiver, 
and those systems with THP at the transmitter and linear equalization at the receiver.
(The linear transceiver is a special case of both systems with the feedback matrix $\B = \0 $; see Figs~\ref{DFE}   and ~\ref{THP1}.)

\subsection{Decision Feedback Equalization}
As shown in Fig.~\ref{DFE},  the transmitted vector is generated by linear precoding, $\x = \P \s$,
and hence the received vector $\y = \H \P \s + \n$. 
The DFE is implemented  using a feedforward matrix $\G^H $ and a strictly lower triangular feedback matrix $\B \in \mathbb{C} $. 
Assuming correct previous decisions, the vector of inputs to the quantizer is  
$\sh = (\G^H \H \P - \B)\s + \G^H \n$.
Defining the error signal $\e = \s - \sh$, and using the assumption $\mathrm{E}_{\s} \{\s \s^H\} = \I$, the mean square error matrix can be written as:
\begin{multline}
\textstyle
\MSE = 
\mathrm{E}_{\s} \{\e \e^H\} =
\C\C^H  - \C \P^H \H^H \G     - \G^H \H \P \C^H  \\
\textstyle
+ \G^H \H \P \P^H \H^H \G     + \G^H \R_n  \G,   \label{DFE_MSE}
\end{multline} 
where $\C = \I + \B$ is a unit diagonal lower triangular matrix. 
The objective is to design the  $\G, \C, \P$ for different design criteria, subject to the transmitter power constraint $\mathrm{E}_{\s} \{\x \x^H\} = \tr (\P \P^H) \le \Ptot$.

\begin{figure}
\centering
\includegraphics[width=0.28\textwidth]{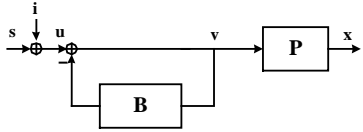}
\caption{ Equivalent linear transmitter model for THP-based system}
\label{THP2}
\end{figure}
\subsection{Tomlinson-Harashima Precoding}
As shown in Fig.~\ref{THP1}, in THP the transmitter performs successive interference pre-subtraction and spatial precoding using the strictly lower triangular matrix $\B  $ and the precoding matrix $\P$, respectively.
We assume that the elements of $\s$  are chosen from a square QAM constellation $\mathcal{S}$ with cardinality $M$ and that $\mathrm{E}_{\s}\{\s \s^H\} = \I$. 
The Voronoi region of this constellation, $\mathcal{V}$, is a square whose side length is $D$. 
Following pre-subtraction of the effect of previously precoded symbols, the transmitter uses the modulo operation so that the symbols of $\v$ lie within the boundaries of $\mathcal{V}$.
The effect of the modulo operation is equivalent to the addition of $\i_k = \i_k^{{re}}   D + \i_k^{{imag}}   D$ to $\s_k$, where $ \i_k^{{re}}, \; \i_k^{{imag}} \in \mathbb{Z}$. 
Using this observation, we obtain the standard linearized model of the transmitter shown in Fig.~\ref{THP2} (e.g.  \cite{F_THP_Book}), in which $ \v = (\I + \B)^{-1} \u = \C^{-1} \u$. 
As a result of the modulo operation, the elements of $\v$ are almost uncorrelated and uniformly distributed over the Voronoi region  $\mathcal{V}$ \cite[Th. 3.1]{F_THP_Book}. 
Therefore, the symbols of $\v$ will have slightly higher average energy than the input symbols $\s$. For a square QAM, we have $\sigv = \mathrm{E}\{|\v_k|^2\} = \frac{M}{M-1} \mathrm{E}\{|\s_k|^2\}$  for all $k$ except the first one \cite{F_THP_Book}. 
For moderate to large values of $M$ this power increase is negligible and the approximation $\mathrm{E}\{\v \v^H\} = \I$ can be used. We will use the  more accurate approximation $\mathrm{E}\{\v \v^H\} = \sigv \I$; e.g., \cite{THP_Simone, F_THP_Book}.
 
For the THP scheme, the received signal vector  can be written as  $\y = \H \P \C^{-1} \u + \n$, 
and hence the receiver's estimate of the of the modified data symbols is 
$     \uh = \G^H \H \P \C^{-1} \u + \G^H \n$.
Following linear equalization, the modulo operation is used to eliminate the effect of the periodic extension of the constellation induced at the transmitter. 
In terms of the modified data symbols, the error signal $\e = \uh - \u = \G^H \H \P \v + \G^H \n - \C \v$ can be used to define the Mean Square Error matrix $\MSE = \mathrm{E}_{\v} \{\e \e^H\}$:
\begin{multline}
\MSE = 
 \sigv \C\C^H  -   \sigv \C \P^H \H^H \G    - \sigv \G^H \H \P \C^H  \\
+\sigv \G^H \H \P \P^H \H^H \G              + \G^H \R_n  \G.   \label{THP_MSE}
\end{multline} 
For the TH precoding model, the transmitter power constraint is given by $\mathrm{E}\{ \x \x^H\} = \sigv \tr( \P\P^H) \le \Ptot$. 

\subsection {General Model}
From equations (\ref{DFE_MSE}) and (\ref{THP_MSE}),  we observe that  the MSE matrix $\MSE$ of both systems has a common form:
\begin{equation}
\MSE = \sig^2 \C\C^H  -  \sig^2  \C \P^H \H^H \G    - \sig^2  \G^H \H \P \C^H   
  +\G^H  \R_y \G, \label{General_MSE}
\end{equation} 
where $\R_y = \sig^2 \H \P \P^H \H^H + \R_n$. 
For the DFE model $\sig^2  = 1 $ while for the TH precoding model $\sig^2  = \sigv$. 
The average transmitter power constraint can be rewritten as $\tr(\P\P^H) \le \Ptot/\sig^2 = P$.

%% file: DFE_THP_design1.tex
\section{optimal feedforward and feedback matrices}   
We consider the joint design of the transceiver matrices $\G, \C, \P$ in order to optimize  system design criteria that are expressed as functions of the MSE of the individual data streams $\E_{ii}$.
We will adopt three-step design approach. 
First, an expression for the  optimal feedforward matrix $\G^H$ will be found as a function of $\C$ and $\P$. 
Second, using the expression of the optimal $\G$, an expression of the optimal $\C$ will be found as a function of $\P$. 
Finally, using the obtained expressions  of $\G$ and $\C$, we will design the optimal precoder $\P$.
\subsection{Optimal feedforward matrix $\G^H$}

For given $\C$ and $\P$, the MSE of the $i^{\text{th}}$ data stream, $\MSEii$, is a convex function of the $i^{\text{th}}$ column of $\G$, denoted $\g_i$, and is independent of other columns. 
Therefore, the columns of $\G$ can be independently optimized to minimize the individual MSEs. 
A similar property was  observed in \cite{Palomar_2003} for linear transceivers. 
Setting the gradient of $\MSEii$ with respect to $\g_i$ to zero, we obtain following expression for the optimal $\G$:
\begin{equation}
\G = \sig^2 \R_y^{-1} \H \P \C^H. \label{opt_G}
\end{equation} 
Since each $\g_i$ independently minimizes the MSE of the $i^{\text{th}}$ data stream, the expression of $\G$ in (\ref{opt_G}) is also the optimal feedforward matrix in the sense of the sum of MSEs, $\tr(\MSE)$. 
Using this expression, the MSE matrix can be written as:
\begin{equation}
\MSE  = \sig^2  \C   ( \I + \sig^2 \P^H \H^H \R_n^{-1} \H \P  )^{-1} \C^H = \C \M \C^H,
\label {MSE_C_F}
\end{equation}
where the matrix inversion lemma has been used.
\subsection{Optimal feedback matrix $\B$}

From (\ref{MSE_C_F}) we observe that the MSE of each data stream $\MSEii$ is convex function of the $i^{\text{th}}$ row of $ \C = \I + \B$ and is independent of the other rows. Therefore, the optimal $\C$ that minimizes the individual MSEs can be obtained by minimizing any convex combination of $\MSEii$. 
By choosing that convex combination to be the sum, our goal reduces to minimizing $\tr(\C \M \C^H)$ subject to $\C$ being unit diagonal lower triangular matrix. 
Using the Cholesky decomposition  $\M = \L \L^H$, where $\L$ is a lower triangular matrix with positive diagonal elements, we can rewrite the objective as $\tr(\C \M \C^H) = \| \C \L \|^2_{F}$, where $\| \cdot \|_{F}$ denotes the Frobenius norm and the product $\C \L$ is positive definite lower triangular matrix \cite{MAtrix_Analysis}. 
Let $\lambda_1(\C \L)\ge \ldots, \ge \lambda_K(\C \L)$ and $\sigma_1(\C \L) \ge  \ldots \ge \sigma_K(\C \L)$ denote the ordered eigen values and singular values, respectively, of the matrix $\C \L$. 
Then the unit diagonal lower triangular $\C$ that minimizes $\tr(\C \M \C^H)$ can be obtained using the following lower bound:
\begin{eqnarray}
\textstyle 
\| \C\L \|^2_{F} = \sum_{i=1}^{K} \sigma_i^2(\C \L)   & \ge &   \sum_{i=1}^{K} \lambda^2_i(\C \L)   \label{step_a}\\
\textstyle
						      &  =  &   \sum_{i=1}^{K}  (\C \L)_{ii}^2 =   \sum_{i=1}^{K} \Lii^2 \label{step_b}, 
\end{eqnarray} 
where the bound in (\ref{step_a}) is obtained by applying Weyl's inequality \cite{Weyl_1949}, and (\ref{step_b}) follows from the fact that $\C \L$ is lower triangular and $\C$ is unit diagonal.
The inequality in (\ref{step_a}) is satisfied with equality when the matrix is normal \cite{Weyl_1949}. 
Since our matrix $\C \L$ is a triangular matrix, it can only be normal if it is diagonal \cite[pp 103]{MAtrix_Analysis}.
Therefore, the matrix $\C$ that attains the lower bound is:
\begin{equation}
\C = \Diag \left( \Lall \right) \L^{-1}.
\end{equation} 
Using this optimal $\C$, the MSE matrix can be  rewritten as:
\begin{equation}
\MSE  = \Diag  \left(  \Lff^2, \ldots, \Lkk^2 \right).
\end{equation} 
We observe that for any given precoding matrix $\P$, the optimal feedforward and feedback  matrices will yield a diagonal MSE matrix, with the individual MSEs being $\MSEii = \Lii^2$.

%% file: DFE_THP_design2.tex
\section {optimal Precoding matrix $\P$}  
Given the optimal   $\G$ and $\C$, the last step is to design a precoding matrix $\P$ to  optimize design criteria expressed as functions of individual MSE of each stream, $\Lii^2$.
We will first derive two inequalities involving $\Lii$ that enable us to characterize the optimal precoder.
\subsection{Preliminaries}
To derive the first inequality, we will use the concept of multiplicative majorization:\newline
\emph{Multiplicative Majorization 
\cite{Weyl_1949, Marshall_1979}:} 
Let $\a, \b \in \mathbb{R}_{+}^{K}$ and let $a_{[1]} \ge \ldots \ge a_{[K]}$ denote the elements of $\a$ in descending order. The vector $\b$ is said to multiplicatively majorize $\a$, $\a \majx \b$, if
$\prod_{i=1}^{j}   \a_{[i]}       \le   \prod_{i=1}^{j}   \b_{[i]}, \text{for} \; j = 1, \ldots, K-1 $ 
and 
$\prod_{i=1}^{K}   \a_{[i]}       =    \prod_{i=1}^{K}   \b_{[i]}$. \newline
An important example of this definition is:
\begin{lem} 
\emph{Weyl \cite{Weyl_1949}:}  
Let $\A \in \mathbb{C}^{K \times K}$ and let $\lambda_i(\A)$ and $\sigma_i(\A)$ denote  the eigen values and singular values of $\A$, respectively. 
Then we have $ (|\lambda_1(\A)|^2,~\ldots,~|\lambda_K(\A)|^2) \majx (\sigma_1^2(\A),~\ldots,~\sigma_K^2(\A)) $.
If $\A$ is normal, then  $|\lambda_i(\A)|= \sigma_i(\A)$.
\label{Weyl_inq}
\end{lem}  
Applying the above lemma to the positive definite lower triangular matrix $\L$, we obtain out first inequality:
\begin{equation}
(\Lff^2, \ldots, \Lkk^2) \majx (\sigma_1^2 (\L), \ldots, \sigma_K^2 (\L)). \label{ineq1_majx}
\end{equation}

The second inequality involves the more  common notation of additive majorization:\newline
\emph{Additive Majorization \cite{Marshall_1979}:} 
Let $\a,~\b \in \mathbb{R}^{K}$. 
The vector $\b$ is said to majorize   $\a$,  $\a~\prec~\b$,  if
$\sum_{i=1}^{j}   \a_{[i]}       \le   \sum_{i=1}^{j}   \b_{[i]},  \text{for} \; j = 1, \ldots, K-1 $ 
and 
$\sum_{i=1}^{K}   \a_{[i]}       =    \sum_{i=1}^{K}   \b_{[i]}$
 
We observe that if elements of $\a$ and $\b$ are positive, then $\a \majx \b \Leftrightarrow \ln(\a) \prec \ln(\b)$. Consequently, (\ref{ineq1_majx}) can be written as:
\begin{equation}
\l \prec \m,
\label{ineq1}
\end{equation} 
where $\l = (\ln \Lff^2, \ldots, \ln \Lkk^2)$ and  $ \m = (\ln \sigma^2_1(\L), \ldots, \ln \sigma^2_K(\L))$. 

To derive the second inequality, we will use the following consequence of  additive majorization:
Any vector $\a~\in~\mathbb{R}^{K}$ majorizes its mean vector $\overline{\a}$ whose elements are all equal to the mean; i.e.,  $\overline{\a}_i~=~\frac{1}{K}~\sum_{i = 1}^{K}~\a_i$. That is, $\overline{\a} \prec \a$. 
Now, since $\M = \L \L^H$, we know that $\prod_{i=1}^{K} \Lii^2 = \det (\L \L^H  ) = \det (\M)$.
As a result, we have $\sum_{i=1}^{K}~\l_i~=~\ln\det(\M)$ and our second inequality is:
\begin{equation}
\lb \prec \l, 
\label{ineq2}
\end{equation} 
where $\lb_i~=~ \frac{1}{K}\ln\det(\M)$. 

The proposed designs will be based on the following classes of functions \cite{Marshall_1979}:
A real-valued function $f(\x)$ defined on a subset $\mathcal{A}$ of $\mathbb{R}^{K}$ is said to be  Schur-convex if 
$\a \prec \b   \:   \textrm{on} \:  \mathcal{A}  \Rightarrow f(\a) \le f(\b)$,
and is said to be Schur-concave if
$\a \prec \b   \:   \textrm{on} \:  \mathcal{A}  \Rightarrow f(\a) \ge f(\b)$.
In particular, we will consider communication objectives that can be expressed as the minimization of a functions of the MSEs of each data stream, $g(\Lff^2, \ldots, \Lkk^2) = g (e^{\l_1}, \ldots, e^{\l_K})) = g(e^{\l})$.

%% file: DFE_THP_design3.tex
\subsection{Schur-convex functions}
Examples of objectives that result in $ g(e^\l) $  being a Schur-convex function of $\l$ include:
minimization of the maximum of  individual MSEs: $g(e^\l) = \max_i e^{\l_i}$; 
minimization of the total MMSE: $g(e^\l) = \sum_i e^{\l_i}$;
and minimization of the (log) determinant MSE matrix:  $\det (\MSE) = \prod_i e^{\l_i}$,
which is also Schur-concave function of $\l$. 
For the DFE model, the SINR of the $i^{\text{th}}$  stream is given by 
$\text{SINR}_i = (1/\text{MSE}_i) - 1 = e^{- \l_i} - 1$. 
Hence, many objectives in terms of SINR and BER can be expressed as Schur-convex functions of $\l$.
As we will show below, the optimal transceiver design is identical for all these objectives. 

If $g(e^\l)$ is a Schur convex function of $\l$, then from $(\ref{ineq2})$ we have that $g(e^{\lb}) \le g(e^\l)$ and the optimal value is obtained when all $\l_i$ are equal to $\l_i = \frac{1}{K} \ln \det (\M)$; i.e., $\E_{ii} = \Lii^2 = \sqrt[K]{\det(\M)}$. 
Since the objective is an increasing function of the individual MSE, the design goal  reduces to minimizing $\det \M$ subject to the power constraint and to the constraint that diagonal elements of the Cholesky  factor of $\M$ are all equal. 
We will start by characterizing the family of solutions that minimize $\det(\M)$ subject to the power constraint, then we will show that there is a member of this family that yields a Cholesky factor of $\M$ with equal diagonal elements. 
Minimizing $\det (\M)$ is equivalent to maximizing the Gaussian mutual information, and  the family of optimal precoders is obtained using a standard water-filling algorithm  \cite{Witsenhausen_1975}. 
In particular, if $\R_H = \sig^2 \H^H \R_n^{-1}\H   = \U \Lambdab_{\H}\U^H$, the family of optimal precoders takes the form: 
\begin{equation}
\P =  \U_1 \hat{{\Phib}} \V = \U_1 [\Phib \quad \0 ] \V, \label{precoder_family}
\end{equation} 
where $ \U_1 \in \mathbb{C}^{N_t \times \Kh}$ contains the eigen vectors of  $\R_{\H}$ corresponding to the $\Kh \le K$ largest eigen values, $\Kh$ and the diagonal positive definite matrix $\Phib$ are obtained from the water- filling algorithm \cite{Witsenhausen_1975}, and $\V \in \mathbb{C}^{K \times K}$ is a unitary matrix degree of freedom. This result shows that for DFE based systems designed according to any Schur-convex function of $\l$, the optimal solution  is information lossless.
To complete the design of $\P$, we need to  select $\V$ such that the Cholesky~decomposition of $\M = \L \L^H$ yields an $\L$ factor with equal diagonal elements. 
Using (\ref{precoder_family}):
\begin{eqnarray}
\M           & =  &   \left(  \V^H (\I + \hat{\Phib}^T  \Lambdab_{\H 1} \hat{\Phib} )^{-1/2}    \right)
	              \left(       (\I + \hat{\Phib}^T  \Lambdab_{\H 1}  \hat{\Phib} )^{-1/2} \V \right) \nonumber \\
	     & =  &  \L \L^H = \R^H \R = (\Q \R)^H(\Q \R),
\end{eqnarray} 
where $\Lambdab_{\H 1}$ is the diagonal matrix containing the largest $\Kh$ eigen values of $\R_{\H}$,
and $\Q$ is a matrix with orthonormal columns. 
Hence, finding $\V$ is equivalent to finding a $\V$ such that QR~decomposition of $(\I + \hat{\Phib}^T  \Lambdab_{\H 1}   \hat{\Phib})^{-1/2} \V $ has an R-factor with equal diagonal.
This problem was solved in \cite{Zhang_2005_QRS} and $\V$  can be obtained by applying the algorithm in \cite{Zhang_2005_QRS} to the matrix $(\I + \hat{\Phib}^T  \Lambdab_{\H 1}   \hat{\Phib} )^{-1/2} $; see also \cite{ Xu_2006_DFE, Jiang_2005_GMD}. 

\subsection{Schur-concave functions}
If $g(e^\l)$ is a Schur-concave function of $\l$, then from ($\ref{ineq1}$) we have $g(e^{\m}) \le g(e^\l)$ and the optimal value is obtained when   $\Lii  =  \sig_i(\L)$. According to Lemma~1, this equality holds when $\L$ is normal matrix. Since $\L$ is a lower triangular matrix,  in order to be normal it must be a diagonal matrix \cite{MAtrix_Analysis}. 
The optimal $\C$ in that case is $\I$. 
That is, in the case of Schur-concave functions of $\l$, the optimal  DFE  design results in linear equalization and optimal TH precoding design results in linear precoding. 
%
Examples of this class of objectives include minimization of product of the MSEs and general (weighted) geometrical mean of MSEs.
%
%

%% file: DFE_THP_sim.tex
\section{Simulation Study}
\begin{figure}
\centering
\includegraphics[width=0.38\textwidth]{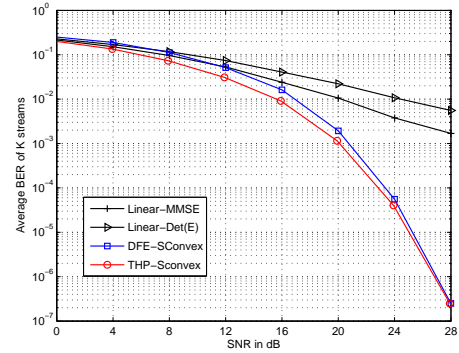}
\caption{BERs of the proposed Schur-convex designs and  the optimal linear transceivers: 
minimum MSE  (Linear-MMSE), 
and maximum mutual information (Linear-Det(E)), 
for  $N_t = N_r= K=4$.}
\label{fig:1}
\end{figure}

We consider a system that transmits $K = 4$ streams of 16-QAM symbols over a $4 \times 4$ slowly fading independent Rayleigh channel with additive white Gaussian noise. 
We plot the average bit error rate (BER)   against the signal to ratio $\Ptot/\tr(\R_n)$.
We compare the performance of the proposed Schur-convex designs for THP and DFE  (which minimize the total MSE among other objectives), with the corresponding linear transceiver design that minimizes total MSE \cite{Palomar_2003, Scaglione_1999a}, and the optimal linear transceiver that maximizes the mutual information (minimizes $\log \det(\MSE)$) \cite{Palomar_2003}. 
The performance advantages of interference cancellation are quite clear from Fig.~\ref{fig:1}. 
%

%% file: DFE_THP_conc.tex
\section{Conclusion}
We developed a unified framework for joint transceiver design of interference (pre-)subtraction schemes for MIMO channels.
We obtained optimal designs for two classes of communication objectives, namely those that are  Schur-convex and  Schur-concave functions of the logarithms of the individual MSEs.
For Schur-convex objectives, the optimal transceiver results in equal individual MSEs. 
For the DFE model, it optimizes both the total MSE and mutual information. 
For the class Schur-concave objectives, the optimal  DFE  design results in linear equalization and the optimal TH precoding design results in linear precoding. 